\documentclass[pra,twocolumn,letterpaper,amsfonts]{revtex4}%

\usepackage[dvips]{graphicx}
\usepackage{amsmath}

\usepackage{amssymb}

\newcommand{\Z}{\mathbb{Z}}

\begin{document}

\title{The kinetic energy operator in the subspaces of wavelet analysis}

\author{J\'anos Pipek}
\affiliation{Department of Theoretical Physics, Institute of Physics,\\
             Budapest University of Technology and Economics,\\
             H--1521 Budapest, Hungary\\}
\author{Szilvia Nagy}
\affiliation{Department of Telecommunication,\\
             Jedlik \'Anyos Institute of Informatics, Electrical and Mechanical Engineering,\\
             Sz\'echenyi Istv\'an University, H--9026 Gy\H{o}r, Egyetem t\'er 1, Hungary\\}


\begin{abstract}

At any resolution level of wavelet expansions the physical observable of the
kinetic energy is represented by an infinite matrix which is ``canonically''
chosen as the projection of the operator $-\Delta/2$ onto the subspace of the
given resolution. It is shown, that this canonical choice is not optimal, as
the regular grid of the basis set introduces an artificial consequence of
periodicity, and it is only a particular member of possible operator
representations. We present an explicit method of preparing a near optimal
kinetic energy matrix which leads to more appropriate results in numerical
wavelet based calculations. This construction works even in those cases, where
the usual definition is unusable (i.e., the derivative of the basis functions
does not exist). It is also shown, that building an effective kinetic energy
matrix is equivalent to the renormalization of the kinetic energy by a momentum
dependent effective mass compensating for artificial periodicity effects.

\end{abstract}

\maketitle

\section{Introduction}

Wavelets are commonly used for analyzing and for a compact storage of complex
distributions like two dimensional images, temporal signals, and even for
solving partial differential equations. Goedecker and Ivanov \cite{goed} solved
the Poisson equation, Cho et al.\ employed wavelets in solving the
Schr\"odinger equation for Hydrogen-like atoms \cite{arias1}. Using this tool,
all electron calculations were also performed within the framework of the local
density approximation applying pseudopotentials and supercells \cite{wei-chou},
Car--Parinello algorithm \cite{tymcz} and in Ref.~\cite{iv-antr} a new approach
for magnetic ordering was presented. Arias and his coworkers developed
Kohn--Sham equations based wavelet method \cite{arias2}, and also tested for
various systems (e.g., \cite{AriasPRB} and \cite{han-cho}).

Wavelet analysis is a popular label applied for the concept of multiresolution
analysis (MRA), which covers a systematically refined basis function set of
Hilbert spaces. We refer to basic textbooks (see e.g., \cite{Daub,Chui}) for
the details.

In our previous works we have shown that the surroundings of a
molecule can be described at a rather rough resolution level
\cite{egy,negy}. We have also demonstrated \cite{ket} that
electron-electron cusp singularity of the two-electron density
operator can be easily reproduced by the method of multiresolution
analysis. In \cite{ot} we have studied the detail structure of the
wave function at various refinement levels using MRA. An adaptive
method was also developed for identifying the fine structure
localization regions, where further refinement of the wave
function is necessary without solving the eigenvalue equation in
the whole subspace expanded by the basis functions of the given
resolution level.

\section{Systematic error in finite resolution eigenfunctions}

While having studied the question, which physical regions of the
potential need a high resolution expansion of the wave function,
we have solved numerically the matrix form of the one particle
Schr\"odinger equation of exactly solvable models, with the
Hamiltonian
\begin{equation}\label{ham}
  H=-\frac{1}{2} \Delta+V(x).
\end{equation}
The algebraic representation of the Schr\"odinger equation
\begin{equation}\label{sch}
  H\Psi_i=E_i\Psi_i
\end{equation}
for the $i$th excited state is derived by considering that
according to the MRA construction, at the resolution level $M$,
the Hilbert space $\mathcal{H}$ is approximated by one of its
subspaces
$\mathcal{H}^{[M]}=\mathrm{span}\{s_{M\ell}(x)|\ell\in\Z\}$, where
the orthonormal basis functions
\begin{equation}\label{smldef}
  s_{M\ell}(x)=2^{M/2} s(2^Mx-\ell)
\end{equation}
are the scaled and translated versions of the ``mother'' scaling
function $s(x)$. The translated scaling functions are ``sitting''
on an equidistant grid of grid length $2^{-M}$. The series of
subspaces $\mathcal{H}^{[M]}$ $(M\to\infty)$ ``approximates'' in a
given sense \cite{Daub} the complete Hilbert space $\mathcal{H}$
and the projectors
\begin{equation}\label{PMproj}
P_M=\sum_{\ell\in\Z}|s_{M\ell}\rangle\langle s_{M\ell}|
\end{equation}
of $\mathcal{H}^{[M]}$ ``approximate'' the identity operator
(although, in the strict mathematical sense the limit of $P_M$
does not exist).

By inserting the approximate identity $P_M$ into (\ref{sch}) one
arrives at
\begin{equation}\label{schPM}
  H P_M\Psi_i\cong E_i\Psi_i.
\end{equation}
Multiplying (\ref{schPM}) by the basis element $s_{Mj}$ from the
left results in
\begin{equation}\label{schmx}
  \sum_{\ell\in\Z} \langle s_{Mj}|H|s_{M\ell}\rangle \,
     \langle s_{M\ell}|\Psi_i\rangle \cong
     E_i\langle s_{Mj}|\Psi_i\rangle.
\end{equation}
How well approximation (\ref{schPM}) works is far from being
understood. Nevertheless, the above method of algebraization is
conventional, and later on, we will refer to this procedure as
``canonical''.

Of course, the canonical method includes the solution of the
eigenvalue problem
\begin{equation}\label{eigcan}
  \sum_{\ell\in\Z} H_{j\ell}^{[M]}
     c_{M\ell}=
     E_i^{[M]} c_{Mj}
\end{equation}
of the Hamiltonian matrix $H_{j\ell}^{[M]}=\langle
s_{Mj}|H|s_{M\ell}\rangle$. The eigenvalue $E_i^{[M]}$ is only an
approximation to the exact eigenvalue $E_i$ (an upper bound for
the ground state), and the eigenvectors $c_{M\ell}$ define an
approximation
\begin{equation}\label{Psiappr}
  \Phi_i^{[M]}(x)=\sum_{\ell\in\Z} c_{M\ell}\; s_{M\ell}(x)
\end{equation}
of the wave function $\Psi_i(x)$. One can not expect, of course,
that (\ref{Psiappr}) gives a better result than the best
approximation
\begin{equation}\label{Psiproj}
 \Psi_i^{[M]}=P_M \Psi_i=\sum_{\ell\in\Z}
 \langle s_{M\ell}|\Psi_i\rangle\, s_{M\ell}
\end{equation}
in the subspace $\mathcal{H}^{[M]}$.

\begin{figure}
  \setlength{\unitlength}{\textwidth}
  \begin{picture}(0.45,0.60)
    \put(0,0.00){\includegraphics[width=0.45\textwidth]{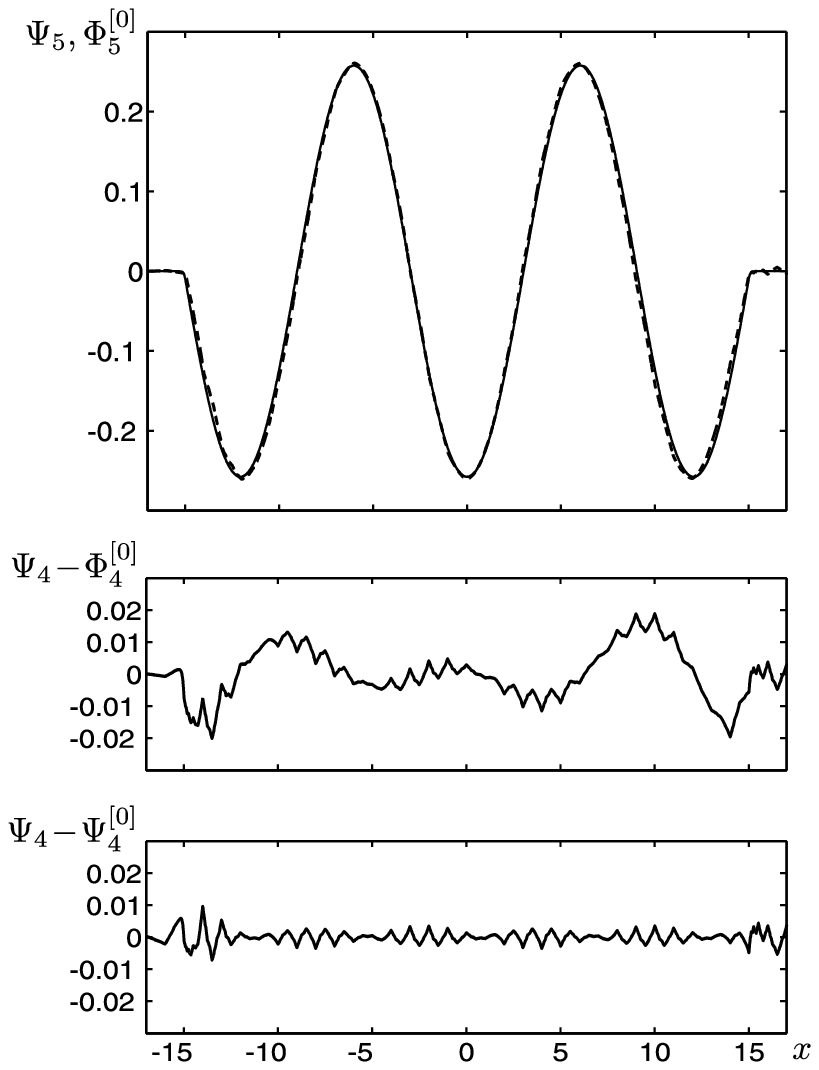}}%
  \end{picture}
\caption{\label{fig:1} Exact (solid line) and approximate (dashed line) wave
functions $\Psi_5$ and $\Phi_5^{[0]}$ of the potential box (\ref{V}), with
$L=15$~a.u., $W=100$~a.u. The difference of the exact wave function and the
solution of the canonical eigenvalue problem is also shown. For a reference the
difference of the exact wave function and its projection to subspace
$\mathcal{H}^{[0]}$ is plotted as well. Atomic units were used.}
\end{figure}

For an illustration, we have chosen the simplest analytically
solvable model of the potential box. The alternative of the free
electron problem was singled out, as the wave functions should be
square integrable in order to be able to successfully describe it
with matrix methods. Fig.~\ref{fig:1} shows the exact excited
state $\Psi_5$, its projection $\Psi_5^{[0]}$ to the subspace of
resolution $M=0$, and the solution $\Phi_5^{[0]}$ related to the
eigenvalue problem (\ref{eigcan}). The Hamiltonian matrix was
calculated using the compactly supported 6 parameter Daubechies
scaling functions \cite{Daub}. As the first derivative of these
basis functions exists, the kinetic energy matrix elements were
determined by
\begin{eqnarray}
  T^{[M]}_{j\ell}&=&\frac{1}{2}\,\langle s_{Mj}|-\Delta| s_{M\ell}\rangle
                  = \frac{1}{2}\,\langle(-i\nabla) s_{Mj}|
                                               (-i\nabla) s_{M\ell}\rangle \nonumber\\
                 &=&\frac{1}{2}\,\int s'_{Mj}(x)s'_{M\ell}(x) dx.
  \label{KinEn}
\end{eqnarray}
Here we have used the fact, that the momentum operator $-i\nabla$
is self adjoint. The potential energy matrix elements were
calculated numerically with the potential function
\begin{equation}\label{V}
  V(x)=\Biggl\{\begin{array}{l@{$\quad \mbox{if }$}l}
                      0 & |x|\leq L,\\
                      W & |x|> L.
              \end{array}\Biggr.
\end{equation}

By a careless analysis of the results one can easily draw
erroneous conclusions. One might think, that at the regions, where
the difference of the exact and approximate solution is large, a
further refinement of the basis set is necessary. Theoretically,
this could be accomplished by adding wavelets sitting in the
regions of large errors. Wavelets are localized basis functions of
the orthogonal complement subspace $\mathcal{W}^{[M]}$ of
$\mathcal{H}^{[M]}$ in the embedding subspace
$\mathcal{H}^{[M+1]}=\mathcal{H}^{[M]}\oplus\mathcal{W}^{[M]}$.

According to Fig.~\ref{fig:1} the large error regions are located
at the steepest parts of the oscillating wave function. At these
places, however, the scaling function expansion can not be of bad
quality, considering that any linear function of the form $ax+b$
can be \emph{exactly} expanded in $\mathcal{H}^{[M]}$ at any
resolution level $M$. As the wave functions at the steepest parts
are almost linear, we do not expect large errors in the scaling
function expansion. This prediction is justified in the third plot
of Fig.~\ref{fig:1}, where the difference of the exact and
projected wave function is shown. The large deviations in the
approximate wave function $\Phi_i^{[M]}$ should have a different
origin.

A careful study of the first part of Fig.~\ref{fig:1} leads to the
conclusion, that the solution of the eigenvalue problem has (apart
from small irregularities) an oscillatory form similar to the
exact solution. The essential difference between the two wave
functions is that the approximate wave function $\Phi_i^{[M]}$ has
a slightly smaller wavelength than the exact one, leading, of
course, to a larger kinetic energy. In the case of the excited
state $i=5$, e.g., $\langle\Psi_5|T|\Psi_5\rangle=0.1352$, while
$\langle\Phi_5^{[0]}|T|\Phi_5^{[0]}\rangle=0.1389$. The same
general experience was gained by studying other excited states of
the box model as well as those of the harmonic oscillator.

In the following sections we will analyze the reasons, why the
above effects appear, and suggest a possible solution.

\section{The kinetic energy operator in the subspaces $\mathcal{H}^{[M]}$}

As in the previous example the systematic error in the wave function occurred
in regions without a potential energy contribution, we conclude, that the
effect is due to the representation of the kinetic energy operator. Posing this
question is not as heretical as one might think for the first sight. In the
original formulation of matrix mechanics by Heisenberg, Born and Jordan, the
physical quantities are represented by infinite matrices satisfying the
appropriate commutation rules. There is no specific prescription for the
determination of their matrix elements. On the other hand, Schr\"odinger works
in the Hilbert space of the square integrable functions $\mathcal{H}$, with a
specific prescription for the operator representation of physical quantities,
in particular, $T=-\Delta/2$. Von Neumann has shown \cite{neumann} the
equivalence of both descriptions with the abstract Hilbert space equipped with
linear operators for physical quantities. One can consider here that the
subspace $\mathcal{H}^{[M]}\subset\mathcal{H}$ is itself an infinite
dimensional separable Hilbert space, and as such, can serve for a complete
description of any quantum mechanical system, on its own right. The significant
difference of the multiresolution expansion from the usual atomic orbital
expansions is that the latter span a finite dimensional subspace, which is in
principle unable to describe a quantum system in all details.

As the subspace $\mathcal{H}^{[M]}$ essentially differs from the
complete space of square integrable functions, it is natural that
the elements of the infinite matrix $K_{j\ell}^{[M]}$
corresponding to the physical quantity of the kinetic energy
differs from the matrix elements $T_{j\ell}^{[M]}=\langle
s_{Mj}|\mbox{$-\Delta/2$}\,|s_{M\ell}\rangle$, since $-\Delta/2$
is the operator representation of the kinetic energy in a
different Hilbert space. However, to avoid any weird
representations, we would like to keep some properties of the
``canonical'' approach, which are summarized below.
\begin{enumerate}
\item Considering, that the scaling functions are usually (but not
necessarily) real, according to (\ref{KinEn})
\begin{equation}\label{TSymm}
  T_{j\ell}^{[M]}=\left(T_{\ell\,j}^{[M]}\right)^*=T_{\ell\,j}^{[M]}.
\end{equation}
\item Using the definition of $s_{M\ell}(x)$ and after a simple
variable transformation in (\ref{KinEn}), one arrives at the shift
invariance property
\begin{equation}\label{Tshinv}
  T_{j\ell}^{[M]}=T_{j-\ell}^{[M]},
\end{equation}
indicating that the kinetic energy is represented by a band
matrix.
\item If the scaling functions are compactly supported on the
interval $[0,N-1]$ (as in the case of Daubechies-$N$ bases),
\begin{eqnarray}
 T_{\ell}^{[M]}&=&\frac{1}{2}\int s'_{M\ell}(x)s'_{M0}(x)dx=0, \nonumber\\
                &&\qquad\qquad \mbox{if } |\ell|>N-2.
 \label{Tsupport}
\end{eqnarray}
Here $N$ is the number of parameters which define the mother
scaling function.
\item Using definition (\ref{smldef}) in (\ref{KinEn}) leads to a
scaling property
\begin{equation}\label{Tscaling}
  T_{j \ell}^{[M]}=2^{2M} T_{j \ell}^{[0]}.
\end{equation}
\item In three spatial dimensions a direct product basis function set
$|j_1 j_2 j_3\rangle = s_{Mj_1}(x_1)s_{Mj_2}(x_2)s_{Mj_3}(x_3)$ is used. As the
Laplacian is a simple sum of second derivatives according to the three spatial
variables, orthonormality of the basis functions results in
\begin{eqnarray}
  \langle j_1 j_2 j_3|-\Delta/2|\ell_1 \ell_2 \ell_3\rangle &=&
  T_{j_1 \ell_1}^{[M]} \delta_{j_2 \ell_2} \delta_{j_3 \ell_3}     \nonumber\\
  &&+ \delta_{j_1 \ell_1} T_{j_2 \ell_2}^{[M]} \delta_{j_3 \ell_3} \nonumber\\
  &&+ \delta_{j_1 \ell_1} \delta_{j_2 \ell_2} T_{j_3 \ell_3}^{[M]}.
  \label{T3D}
\end{eqnarray}
Consequently, the case of the 3D kinetic energy operator is straightforwardly
reduced to the one dimensional representation.
\end{enumerate}
Hermiticity (\ref{TSymm}) and translational invariance
(\ref{Tshinv}) are natural requirements for any operator
representations of the kinetic energy. Though we are not obliged
to keep property (\ref{Tsupport}), this is one of the most
attractive features of using compactly supported basis sets, thus
this prescription is applied as well. Transformation of the
kinetic energy matrix elements with increasing resolution (like
the scaling property (\ref{Tscaling})) will be discussed later.
For the matrix $K^{[M]}$ representing the kinetic energy we
require the followings at any resolution level $M$
\begin{eqnarray}\label{KSymm}
  K_{j\ell}^{[M]}&=&\left(K_{\ell\,j}^{[M]}\right)^*=K_{\ell\,j}^{[M]},\\
\label{Kshinv}
  K_{j\ell}^{[M]}&=&K_{j-\ell}^{[M]}=K_{|j-\ell|}^{[M]},\\
\label{Ksupport}
 K_{\ell}^{[M]}&=&0, \qquad \mbox{if
 } |\ell|>N-2.
\end{eqnarray}
In three dimensions we additionally apply
\begin{eqnarray}
  K^{[M]}_{j_1 j_2 j_3,\ell_1 \ell_2 \ell_3} &=&
  K_{j_1 \ell_1}^{[M]} \delta_{j_2 \ell_2} \delta_{j_3 \ell_3}     \nonumber\\
  &&+ \delta_{j_1 \ell_1} K_{j_2 \ell_2}^{[M]} \delta_{j_3 \ell_3} \nonumber\\
  &&+ \delta_{j_1 \ell_1} \delta_{j_2 \ell_2} K_{j_3 \ell_3}^{[M]}.
  \label{K3D}
\end{eqnarray}

The role of operators assigned to observables is to provide the
possible and expectation values of the corresponding physical
quantities. A proper representation of the kinetic energy should
give the known $k^2/2$ eigenvalues with eigenvectors which give
reasonable approximations of the free electron wave function
$e^{ikx}$. In the following considerations we will prove that the
best approximation $P_M e^{ikx}$ is really an eigenvector of any
matrix satisfying the requirements
(\ref{KSymm})--(\ref{Ksupport}). The question remains to clarify
how well the equality
\begin{equation}\label{eigeq?}
  \sum_{\ell\in\Z} K_{j\ell}^{[M]} \langle
  s_{M\ell}|e^{ikx}\rangle\ \  \raisebox{1.2ex}{?}\!\!\!\!\!=\
  \frac{k^2}{2} \langle s_{M j}|e^{ikx}\rangle
\end{equation}
is satisfied.

Defining the Fourier transform by $\hat{f}(\xi)=(2\pi)^{-1/2} \int
f(x)e^{-i\xi x}dx$ and using definition (\ref{smldef}) we have
\begin{equation}\label{expproj}
 \langle s_{M\ell}|e^{ikx}\rangle = 2^{-M/2}
     e^{ik_M\ell}(2\pi)^{1/2}\; \hat{s}(-k_M),
\end{equation}
where $k_M=2^{-M}k$ is the scaled wave number. Using this
expression in the left hand side of (\ref{eigeq?})
\begin{eqnarray}\label{eigeq?2}
  &&\sum_{\ell\in\Z} K_{j\ell}^{[M]} \langle
  s_{M\ell}|e^{ikx}\rangle=                                          \nonumber\\
  &&\sum_{\ell\in\Z} K_{j\ell}^{[M]}\, e^{ik_M(\ell-j)}
  2^{-M/2} e^{ik_M j}(2\pi)^{1/2}\; \hat{s}(-k_M)=                   \nonumber\\
  &&\qquad\left(\sum_{\ell\in\Z}K_{j-\ell}^{[M]}\; e^{ik_M(\ell-j)}\right)
  \langle s_{Mj}|e^{ikx}\rangle.
\end{eqnarray}
It is clear that just the shift invariance (\ref{Kshinv}) ensures
that the projection of the free electron wave function to
$\mathcal{H}^{[M]}$ is an eigenfunction of $K^{[M]}$, with the
eigenvalue
\begin{equation}\label{eigenval}
  \varepsilon^{[M]}(k)= \sum_{\ell\in\Z} K_\ell^{[M]} e^{-ik_M\ell}.
\end{equation}
As according to (\ref{KSymm}) $K_\ell^{[M]}$ is Hermitian, its
eigenvalues $\varepsilon^{[M]}(k)$ are real and
\begin{equation}\label{epssymm}
 \varepsilon^{[M]}(-k)=\sum_{\ell\in\Z} K_\ell^{[M]} e^{ik_M\ell}=
 \sum_{\ell\in\Z} K_{-\ell}^{[M]} e^{-ik_M\ell}=\varepsilon^{[M]}(k),
\end{equation}
i.e., $\varepsilon^{[M]}(k)$ is symmetric in $k$. The above
natural physical requirements are satisfied for all reasonably
chosen $K^{[M]}$.

The fundamental question is how well $\varepsilon^{[M]}(k)$ approximates the
free electron kinetic energy $k^2/2$. In order to understand this, some
properties of $\varepsilon^{[M]}(k)$ will be studied below. As the argument
$k_M$ in definition (\ref{eigenval}) exponentially decreases with increasing
resolution $M$, it is natural to consider the Taylor expansion of
$\varepsilon^{[M]}(k)$. Due to the symmetry of $\varepsilon^{[M]}(k)$ all its
odd order derivatives should be zero in $k=0$. This condition is equivalent to
$\sum_{\ell\in\Z} \ell^{2n+1} K_\ell^{[M]}=0$ for any $n=0,1,\ldots$, which
follows from (\ref{Kshinv}). In the ideal case $\varepsilon^{[M]}(k)$ would be
equal to $k^2/2$ requiring
\begin{eqnarray}\label{Taylor1}
  \varepsilon^{[M]}(0)=\sum_{\ell\in\Z}K_\ell^{[M]}&=&0,\\
     \label{Taylor2}
  \frac{d^2 \varepsilon^{[M]}}{dk^2} (0)=
  -2^{-2M} \sum_{\ell\in\Z}\ell^2 K_\ell^{[M]}&=&1,\\
     \label{Taylor3+}
  \frac{d^{2n} \varepsilon^{[M]}}{dk^{2n}} (0)=
  (-1)^n2^{-2nM} \sum_{\ell\in\Z}\ell^{2n} K_\ell^{[M]}&=&0,
\end{eqnarray}
for $n\geq2$. In Appendix~\ref{sec:appA}, we have shown that for
the canonical kinetic energy matrix sum rules (\ref{sumrule1}) and
(\ref{sumrule3}) ensure that (\ref{Taylor1}) and (\ref{Taylor2})
are satisfied. It is easy to see, however, that regardless of the
choice of $K_\ell^{[M]}$ the equivalence of $\varepsilon^{[M]}(k)$
and $k^2/2$ can never be perfect. According to (\ref{Ksupport})
$\varepsilon^{[M]}(k)$ is a finite trigonometric polynomial,
consequently it is $2\pi$ periodic, and can not coincide with the
free electron energy in the whole range $-\infty<k<\infty$.
\begin{figure}
  \setlength{\unitlength}{\textwidth}
  \begin{picture}(0.45,0.45)
    \put(0,0.00){\includegraphics[width=0.45\textwidth]{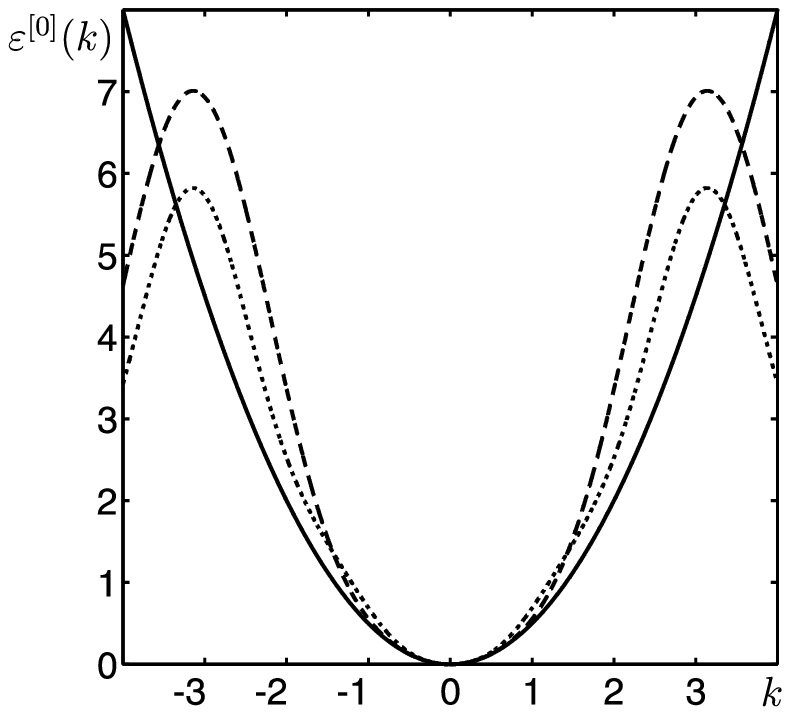}}%
  \end{picture}
\caption{\label{fig:2} The canonical kinetic energy function
$\varepsilon_{\mathrm{can}}^{[0]}(k)$ (dashed line) of the Daubechies-6 basis
set compared to the free particle kinetic energy $k^2/2$ (solid line). The
kinetic energy $\varepsilon_{\mathcal{F}}^{[0]}(k)$ of the Fourier type
approximation with the same number of matrix elements is plotted by a dotted
line. Atomic units were used.}
\end{figure}

Fig.~\ref{fig:2} shows that the canonical kinetic energy function
approaches the ``Brillouin zone'' boundaries at $k=-\pi$ and
$k=\pi$ with a horizontal tangent. This difficulty can never be
resolved by choosing any set of $K_\ell^{[M]}$ values for $2\pi$
periodic $\varepsilon^{[M]}(k)$ functions, which leads to the
conclusion, that any multiresolution quantum mechanics loses its
applicability for the energies with $|k|=2\pi/\lambda\approx\pi$.
In other words, for the wavelength of the particle we get the
condition $\lambda/2\ll 1$. It is not surprising from the physical
point of view, since 1 is the grid length of the scaling function
basis set at resolution level $M=0$ and no wave functions with a
wavelength comparable to the grid length are expected to be
described in a satisfactory manner. If the resolution increases,
however, the scaling property (\ref{Tscaling}) results in
\begin{equation}\label{epsilonscaling}
  \varepsilon^{[M]}_\mathrm{can}(k)=2^{2M} \sum_{\ell\in\Z}
     T_\ell^{[0]}e^{-ik_M\ell}=
     2^{2M}\varepsilon^{[0]}_\mathrm{can}(2^{-M}k).
\end{equation}
Since the argument of $\varepsilon^{[0]}_\mathrm{can}$ decreases
exponentially with increasing resolution, the quality of the
Taylor expansion becomes increasingly better, as it can be traced
in Fig.~\ref{fig:2}, in the close neighborhood of $k=0$. Although
it is very satisfying that the canonical kinetic energy can
reproduce the exact values in the infinite resolution limit, in a
practical calculation, however, one should stay at a relatively
low resolution $M$. At these resolutions the function
$\varepsilon^{[M]}_\mathrm{can}(k)$ performs rather poorly for
larger $k$, and there is a reasonable hope to find matrix elements
$K_\ell^{[M]}$ which provide significantly better approximation of
the function $k^2/2$.

\section{Assignment strategies for the optimal kinetic energy matrix}

There are several possible decisions for choosing the matrix
elements $K_\ell^{[M]}$ in order to approximate $k^2/2$ by a
finite trigonometrical polynomial expansion of the form
(\ref{eigenval}). As a first remark, we recall the scaling
property (\ref{Tscaling}) of the canonical kinetic energy matrix.
Since the values of $K_\ell^{[M]}$ are not determined by a
mathematical formula (similar to (\ref{KinEn})), their scaling
behavior can not be derived. On the other hand, however, the
necessary requirement (\ref{Taylor2}) implies that in the leading
order the kinetic energy matrix elements should scale as
$K_\ell^{[M]}\sim 2^{2M}$. We would like to emphasize, that
choosing the scaling formula
\begin{equation}\label{Kscaling}
  K_\ell^{[M]}=2^{2M} K_\ell^{[0]}
\end{equation}
is by no means a must, we have decided to apply it in order to
decrease the many degrees of freedom of the problem. After this,
we still have to determine the values of the zeroth level matrix
elements, for which we consider two different philosophies.

\subsection{The Fourier series approach}

The best possibility we can expect using an expression of the form
(\ref{eigenval}) is that the function $k^2/2$ is correctly
described in the interval $(-\pi,\pi)$. This, however, requires
the infinite Fourier series expansion
\begin{equation}\label{Fourierk2}
 \frac{k^2}{2}= \frac{\pi^2}{6}+
 2\sum_{\ell=1}^{\infty} \frac{(-1)^\ell }{\ell^2} \cos(\ell k).
\end{equation}
Applying (\ref{Ksupport}) leads to a truncation of
(\ref{Fourierk2}) at $\ell=N-2$. It is clear that identifying
$K_\ell^{[0]}$ with the expansion coefficients of the truncated
series  would not satisfy any of the criteria
(\ref{Taylor1})--(\ref{Taylor3+}). It is an elementary requirement
that a particle with zero momentum should have zero kinetic energy
(criterion (\ref{Taylor1})). On the other hand, in the large
resolution limit we should recover the exact kinetic energy, and
as we have discussed above, this is equivalent to (\ref{Taylor2}).
Considering these arguments, we suggest the following truncation
process
\begin{equation}\label{KFuorier}
 \alpha\mathcal{F}_\ell^{N}= \left\{\begin{array}{l@{\ \mbox{for}\ }l}
      \!\frac{\pi^2}{6} & \ell=0,\\
      \!\frac{(-1)^\ell}{\ell^2} & 1\leq|\ell|\leq N-3,\\
      \!-\frac{1}{2} \left(\frac{\pi^2}{6}+
              2\sum_{\ell=1}^{N-3} \frac{(-1)^\ell}{\ell^2}
              \right) & |\ell|=N-2.
              \end{array}\right.
\end{equation}
The definition $K_\ell^{[M]}=2^{2M}\mathcal{F}_\ell^{N}$
automatically satisfies sum rule (\ref{Taylor1}), whereas with an
appropriate choice of the normalization factor $\alpha$,
(\ref{Taylor2}) can also be fulfilled. The kinetic energy function
$\varepsilon^{[0]}_\mathcal{F}(k)$ calculated according to
(\ref{eigenval}) using the matrix elements determined by
(\ref{KFuorier}) is plotted in Fig.~\ref{fig:2}. According to the
figure, the Fourier type approach results in a weaker quality
approximation than the canonical calculation, especially in the
low energy region. This effect can be traced in Tab.~\ref{tab:1};
both the total energy and the wave function deviate more from the
exact quantities than those of the canonical calculation, except
for the wave function of higher excited states.
\begin{table*}
\caption{\label{tab:1} The error of the total energy $E_i^{[0]\mathcal{F}}$ and
the wave function $\Phi_i^{[0]\mathcal{F}}$ of the Fourier method compared to
the ones of the canonical quantities $E_i^{[0]}$ and $\Phi_i^{[0]}$ for the
Daubechies-6 basis set. $E_i$ and $\Psi_i$ are the exact total energy and wave
function, respectively. Label $i=1$ indicates the ground state, whereas
$i=2,3,4,5$ are the successive excited states of the potential box (\ref{V}),
with $L=15$~a.u., $W=100$~a.u.}
\bigskip
\begin{tabular}{c@{\qquad}c@{\qquad}c@{\qquad}c@{\qquad}c}
     \hline\hline
     $i$ & $|E_i-E_i^{[0]}|$ &  $|E_i-E_i^{[0]\mathcal{F}}|$ & $\|\Psi_i-\Phi_i^{[0]}\|$ & $\|\Psi_i-\Phi_i^{[0]\mathcal{F}}\|$  \\
     \hline
 1 & $ 1.0802\times 10^{-4}$ & $2.7721\times 10^{-4}$ & $1.0392\times 10^{-2}$ & $1.4545\times 10^{-2}$\\
 2 & $ 4.4247\times 10^{-4}$ & $1.6186\times 10^{-3}$ & $1.8878\times 10^{-2}$ & $2.1309\times 10^{-2}$\\
 3 & $ 1.0681\times 10^{-3}$ & $5.4125\times 10^{-3}$ & $2.7654\times 10^{-2}$ & $2.6331\times 10^{-2}$\\
 4 & $ 2.1982\times 10^{-3}$ & $1.3550\times 10^{-2}$ & $3.6618\times 10^{-2}$ & $2.8380\times 10^{-2}$\\
 5 & $ 4.3310\times 10^{-3}$ & $2.7925\times 10^{-2}$ & $4.5968\times 10^{-2}$ & $2.7997\times 10^{-2}$\\
     \hline\hline
\end{tabular}
\end{table*}

\subsection{The Taylor series approach}

As we have seen, the quality of the $\varepsilon^{[0]}(k)$ for
smaller $k$ values is essential both in rough resolutions and also
in the limit $M\to \infty$. This leads to the conclusion that the
Taylor expansion of the kinetic energy function should satisfy as
much conditions of (\ref{Taylor1})--(\ref{Taylor3+}) as possible
with the given number of non-zero matrix elements $K_\ell^{[M]}$.
We suggest the following scheme for determining the optimal
kinetic energy matrix. According to (\ref{Kshinv}) and
(\ref{Ksupport}) the number of non-zero, essentially different
matrix elements is $N-1$, offering too much freedom in the
optimization process. Consequently, we have decided to keep only
one independent parameter $t$ and to define the kinetic energy
matrix elements by
\begin{equation}\label{kinmxTaylor}
  K_\ell^{[M]}=2^{2M}\mathcal{T}_\ell^{N}(t)
\end{equation}
with
\begin{equation}\label{KTaylor}
 \alpha\mathcal{T}_\ell^{N}(t)= \left\{\begin{array}{l@{\quad\mbox{for}\quad}l}
      1      & \ell=0,\\
      t      & |\ell| =1,\\
      t_\ell(t) & 2\leq|\ell|\leq N-2.
              \end{array}\right.
\end{equation}
Quantities $t_\ell(t)$ are determined by the solution of the
linear system of equations
\begin{eqnarray}\label{tmxeq}
  \left(\begin{array}{cccc}
  1         &   1       & \ldots & 1      \\
  2^4       & 3^4       & \ldots & (N-2)^4\\
  2^6       & 3^6       & \ldots & (N-2)^6\\
  \vdots    & \vdots    & \ddots & \vdots \\
  2^{2(N-3)}& 3^{2(N-3)}& \ldots & (N-2)^{2(N-3)}
  \end{array}\right)
  \left(\begin{array}{c}
  t_2\\
  t_3\\
  \vdots\\
  t_{N-2}
  \end{array}\right)&=&             \nonumber\\
  \left(\begin{array}{c}
  -1/2-t\\
  -t\\
  \vdots\\
  -t
  \end{array}\right). \quad &&
\end{eqnarray}
The normalization factor
\begin{equation}\label{alphanorm}
  \alpha=-2t-2\sum_{\ell=2}^{N-2} \ell^2 t_\ell(t).
\end{equation}
It is easy to verify that the conditions (\ref{Taylor1}) and
(\ref{Taylor2}) are always satisfied by these values and
(\ref{Taylor3+}) fulfills until $n=N-3$.

\begin{figure}[b]
  \setlength{\unitlength}{\textwidth}
  \begin{picture}(0.45,0.45)
    \put(0,0.00){\includegraphics[width=0.45\textwidth]{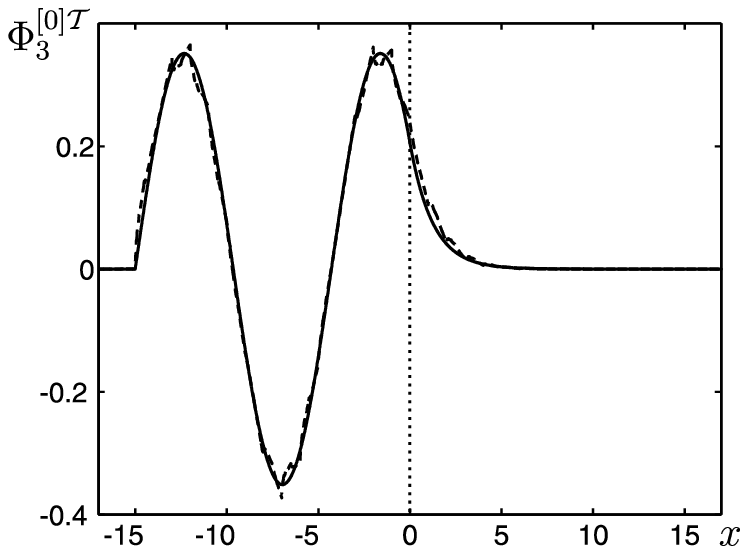}}%
  \end{picture}
\caption{\label{fig:3} Exact (solid line) and approximate (dashed line) wave
functions $\Psi_1$ and $\Phi_1^{[0]\mathcal{T}}$ of the potential model
(\ref{Vstep}), with $L=15$~a.u., $W=100$~a.u. and $V_0=0.5$~a.u., using the
Daubechies-4 scaling function in the Taylor series method. Atomic units were
applied.}
\end{figure}

Notice, that equations (\ref{tmxeq}) have a solution even in the
case of $N=4$, i.e., for the Daubechies-4 scaling functions, where
the canonical kinetic energy matrix elements are not defined at
all, as the application of formula (\ref{KinEn}) requires the
derivative of the scaling function, which does not exist in this
case. For an illustration we have carried out a calculation for
the potential
\begin{equation}\label{Vstep}
  V(x)=\left\{\begin{array}{l@{$\quad \mbox{if }$}l}
                      0   & -L\leq x< 0,\\
                      V_0 & 0\leq x\leq L,\\
                      W   & |x|> L,
              \end{array}\right.
\end{equation}
with Daubechies-4 scaling functions and with the Taylor series
based method outlined above. The optimal value of the parameter
$t$ was determined as described later. Fig.~\ref{fig:3} shows the
exact and the $M=0$ level approximated wave function of the 2nd
excited state of the model (\ref{Vstep}). This state was selected
in order to demonstrate the capabilities of the method, not only
in the free electron case but in the classically unavailable
regions (right hand side of the box, $E_i<V_0$), as well.
It is seen that the construction of the kinetic energy matrix by the Taylor
expansion method leads to a rather satisfactory result even at this low
resolution approximation.

\begin{figure}
  \setlength{\unitlength}{\textwidth}
  \begin{picture}(0.45,0.60)
    \put(0,0.00){\includegraphics[width=0.45\textwidth]{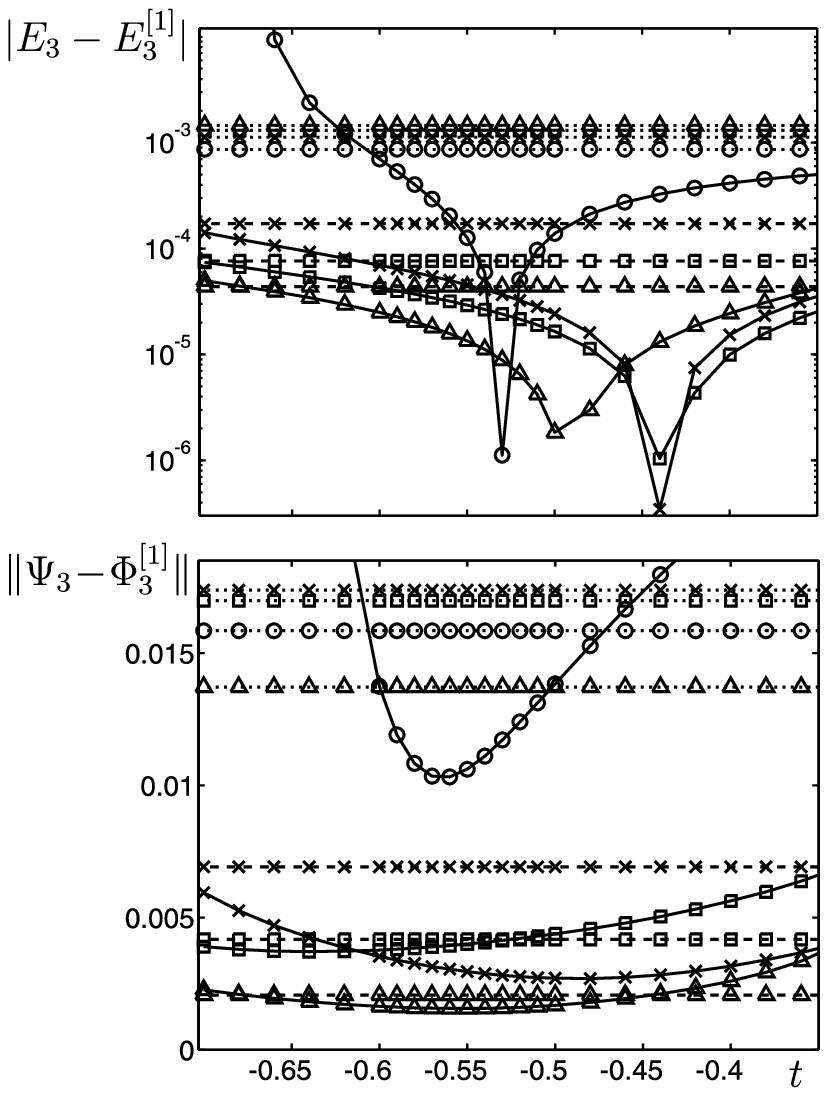}}%
  \end{picture}
\caption{\label{fig:4} The deviation of the approximate total energy
$E_1^{[0]\mathcal{T}}$ and wave function $\Phi_1^{[0]\mathcal{T}}$ from their
exact counterparts $E_1$ and $\Psi_1$ (solid lines) of the potential model
(\ref{Vstep}), with $L=17$~a.u., $W=10$~a.u. and $V_0=0.01$~a.u., using the
Taylor series method. As references, horizontal lines indicate the values of
the canonical (dashed line) and the Fourier series method (dotted line)
results. Sign $\bigcirc$ stands for the results of the Daubechies-4 scaling
function, $\times$, $\square$ and $\triangle$ mean Daubechies-6, Daubechies-8
and Daubechies-10 calculations, respectively. Atomic units were applied.}
\end{figure}

For finding the optimal values of $t$ in the construction
(\ref{KTaylor})--(\ref{alphanorm}) we have plotted the deviation
of the approximate total energy from the exact one as a function
of the parameter $t$ in Fig.~\ref{fig:4}. Similarly, the norm of
the difference of the exact and approximated wave function of the
model potential (\ref{Vstep}) is also shown.
It can be realized, that in quite a broad range of the parameter $t$, the
kinetic energy matrix of the Taylor's expansion method leads to a considerably
better quality result than the canonical choice, regarding both the total
energy and the wave function. The Fourier series method, however, leads to
unacceptable solutions.

According to Fig.~\ref{fig:4}, a minimum in the error functions
can be identified which defines the optimum value of parameter
$t$. We would like to emphasize, that around the minima the curves
are flat, even for the energy deviations (consider, the
logarithmic scale on the vertical axis). The error curves are
similar to all of the lower energy excited states, and even the
positions of the minima are almost the same. As the worst example,
we adduce the case of Daubechies 8 scaling function where the
average value of the optimal $t$ calculated for the energy and
wave function error curves of several excited states is 0.57, with
a standard deviation of $\pm 0.06$. Also, by playing around the
parameters $L$, $W$ and $V_0$ of model (\ref{Vstep}) we could
realize, that the optimum $t$ values are essentially potential
independent. Extending the scope of the investigations to finer
resolution levels $M=0,1,2,3,4$ we have found that the position of
the minima of the error curves stabilizes at $M=1$, there is no
significant change at fine resolutions. Further, we have studied
the exactly solvable model of the harmonic oscillator with the
potential
\begin{equation}\label{Vosc}
  V(x)=\frac{\omega^2}{2}x^2.
\end{equation}
We have experienced the same behavior as in the case of the
previous physical system (\ref{Vstep}). The positions of the
optimum $t$ values are independent of the excitation level $i$, of
the resolution $M$ and also of the value of the potential width
$\omega^{-1}$. Finally, we conclude, that to a good approximation,
the optimum values of parameter $t$ in the kinetic energy matrix
definition (\ref{kinmxTaylor})--(\ref{alphanorm}) can be chosen
model independently, which works also for many low lying excited
states. In Tab.~\ref{tab:2} we summarize our recommendations for
the the best values of $t$ for various Daubechies basis function
sets.
\begin{table}
\caption{\label{tab:2} The recommended kinetic energy matrix parameter $t$ for
the Taylor series based construction (\ref{kinmxTaylor})--(\ref{alphanorm}) in
case of Daubechies-$N$ scaling function sets ($N=4,6,8,10$).}
\bigskip
\begin{tabular}{c@{\qquad}c@{\qquad}c@{\qquad}c@{\qquad}c}
     \hline\hline
 $N$ & 4 &  6 & 8 & 10 \\
     \hline
 $t$ & $-0.54$ &  $-0.47$ & $-0.57$ & $-0.58$ \\
     \hline\hline
\end{tabular}
\end{table}
As the error curves are rather flat around the minima, the values in the table
can be changed by $\pm 10$ percent without a considerable change in the quality
of the total energy and wave function approximations.

\section{Summary}

Studying exactly solvable models in the framework of multiresolution analysis
we have found that the representation of the kinetic energy operator plays an
essential role in the quality of the results achieved by approximate solutions
at a given resolution level $M$. The regular grid of the scaling function basis
set introduces an artificial consequence of periodicity. Instead of a free
particle, the MRA expansion describes rather an electron with a momentum
dependent effective mass $m^*(k)=k^2/(2\varepsilon^{[M]}(k))$, where the
function $\varepsilon^{[M]}(k)$ is determined by the matrix elements of the
kinetic energy matrix. We have shown that in the case of resolution level $M=0$
the MRA expansion loses its applicability if the kinetic energy approaches or
exceeds the value $E_\mathrm{kin}=\pi^2/2$ a.u.\ (corresponding to the limit
value $k=\pi$). By increasing the resolution, the applicability range extends
exponentially as $k<2^M\pi$, $E_\mathrm{kin}<2^{2M}\pi^2/2$ a.u., and
artificial periodicity effects disappear in this limit.

However, in the numerical practice the level of resolution should
be kept as low as possible, in order to avoid the need for
extensive computational resources. We have demonstrated that for
low resolutions the kinetic energy of the numerical calculations
is overestimated compared to the exact values. The effect is due
to the fact that the free particle energy $k^2/2$ is improperly
reproduced by its $2\pi$-periodic approximation
$\varepsilon_\mathrm{can}^{[M]}(k)$. This reproduction can
considerably be improved by introducing alternative matrix
elements of the kinetic energy matrix instead of the canonical
ones. Both the total energy and wave function improvements are
well pronounced at low resolutions as it can be traced from
Fig.~\ref{fig:4}. A close optimal, system and eigenstate
independent choice of the kinetic energy matrix elements is
derived from the formulas of the Taylor series approach
(\ref{kinmxTaylor})--(\ref{alphanorm}) and from our suggestion for
its parameter value $t$ in Tab.~\ref{tab:2}. At an arbitrary
resolution level $M$ the kinetic energy matrix elements are
calculated as
\[
  K_{j\ell}^{[M]}=K_{|j-\ell|}^{[M]}=2^{2M}\mathcal{T}_{|j-\ell|}^{N},
\]
where $N$ determines the number of essential matrix elements, as
$\mathcal{T}_{|j-\ell|}^{N}=0$ if $|j-\ell|>N-2$. Tab.~\ref{tab:3}
lists the values of $\mathcal{T}_\ell^N$ for various
Daubechies-$N$ scaling function sets.
\begin{table*}
\caption{\label{tab:3} The recommended kinetic energy matrix elements for the
Taylor series based construction in case of Daubechies-$N$ scaling function
sets.}
\bigskip
\begin{tabular}{c@{\qquad}r@{\qquad}r@{\qquad}r@{\qquad}r}
     \hline\hline
                   &       $N=4\quad\ $ &      $N=6\quad\ $ &     $N=8\quad\ $ &  $N=10\quad\ $ \\
     \hline
 $\mathcal{T}_0^N$ &    $1.3157894737$  &   $1.0269360269$  &  $1.4668325041$  &  $1.5177613012$\\
 $\mathcal{T}_1^N$ &   $-0.7105263158$  &  $-0.4826599327$  & $-0.8360945274$  & $-0.8803015547$\\
 $\mathcal{T}_2^N$ &    $0.0526315789$  &  $-0.0586700337$  & $ 0.1207733653$  & $ 0.1495444216$\\
 $\mathcal{T}_3^N$ &                    &  $ 0.0326358826$  & $-0.0206082682$  & $-0.0344316374$\\
 $\mathcal{T}_4^N$ &                    &  $ 0.0047739298$  & $ 0.0027102582$  & $ 0.0074722170$\\
 $\mathcal{T}_5^N$ &                    &                   & $-0.0002005664$  & $-0.0013201227$\\
 $\mathcal{T}_6^N$ &                    &                   &  $0.0000034864$  & $ 0.0001689233$\\
 $\mathcal{T}_7^N$ &                    &                   &                  & $-0.0000133612$\\
 $\mathcal{T}_8^N$ &                    &                   &                  & $ 0.0000004634$\\

     \hline\hline
\end{tabular}
\end{table*}
In three spatial dimensions expression (\ref{K3D}) should be applied. With the
suggested method it is possible to define a kinetic energy matrix even in the
case of the Daubechies-4 basis set, where the scaling function is not
differentiable, thus the canonical approach is not applicable.

\section*{ACKNOWLEDGMENTS}

This work was supported by the Orsz\'agos Tudom\'anyos Kutat\'asi
Alap (OTKA), Grant Nos. T046868 and NDF45172.

\appendix
\section{Some elementary properties of the canonical kinetic
energy matrix}\label{sec:appA}

We will prove here, that the canonical kinetic energy matrix
elements defined by (\ref{KinEn}) satisfy simple sum rules as
\begin{eqnarray}\label{sumrule1}
 \sum_{\ell\in\Z} T_\ell^{[M]}&=&0, \\
 \label{sumrule2}
 \sum_{\ell\in\Z} \ell\, T_\ell^{[M]}&=&0,\\
 \label{sumrule3}
 \sum_{\ell\in\Z} \ell^2 T_\ell^{[M]}&=&-2^{2M}
\end{eqnarray}
with the notation introduced in (\ref{Tshinv}). Of course, these
relations hold only if the canonical kinetic energy matrix exists,
i.e., if the scaling function is differentiable. This condition is
satisfied for the Daubechies basis sets with 6 or more parameters.

\noindent%
\emph{Proof of (\ref{sumrule1}).} At any resolution level $M$ the
scaling function basis set is capable to exactly expand any
constant function \cite{Daub,Chui}, consequently, for any $x$
\begin{equation}\label{partunity}
  \sum_{\ell\in\Z} c^{[M]}_\ell s_{M\ell}(x) =1
\end{equation}
with the expansion coefficients $c^{[M]}_\ell=2^{-M/2}$.
Differentiating, multiplying by $s'_{M0}(x)/2$ and integrating one
gets
\begin{equation}\label{proof1}
   \sum_{\ell\in\Z}\frac{1}{2}\int s'_{M0}(x)s'_{M\ell}(x) dx =\sum_{\ell\in\Z}
   T_\ell^{[M]}=0.
\end{equation}

\noindent%
\emph{Proof of (\ref{sumrule2}).} The basis set
$\{s_{M\ell}|\ell\in\Z\}$ exactly expands the identity function,
i.e., for all $x$
\begin{equation}\label{xexpansion}
  \sum_{\ell\in\Z} c^{[M]}_\ell s_{M\ell}(x) =x
\end{equation}
with the appropriate expansion coefficients
\begin{equation}\label{xexpansioncoeff}
c^{[M]}_\ell=\int x s_{M\ell}(x) dx=
 2^{-3M/2}(\mu_1+\ell),
\end{equation}
where we have applied definition (\ref{smldef}), a proper integral
variable transformation, and the fact that $\int s(y)dy=1$
\cite{Daub}. The quantity $\mu_1=\int y s(y)dy$ is the first
momentum of the mother scaling function. Differentiating
(\ref{xexpansion}), multiplying by $s'_{M0}(x)/2$ and integrating
we arrive at
\begin{eqnarray}\label{proof21}
 &&\sum_{\ell\in\Z}c_\ell^{[M]}\frac{1}{2}\int s'_{M0}(x)s'_{M\ell}(x) dx =
   \sum_{\ell\in\Z} c_\ell^{[M]} T_\ell^{[M]}=                         \nonumber\\
 &&\qquad
   \frac{1}{2} \int s'_{M0}(x)dx=
   \frac{1}{2} \left[s_{M0}(x)\right]_{-\infty}^\infty.
\end{eqnarray}
As the scaling functions are square integrable,
$s_{M0}(\pm\infty)=0$, and
\begin{eqnarray}\label{proof22}
  0&=&\sum_{\ell\in\Z} c_\ell^{[M]}T_\ell^{[M]}=                       \nonumber\\
  &&2^{-3M/2}\mu_1 \sum_{\ell\in\Z} T_\ell^{[M]}+
    2^{-3M/2}\sum_{\ell\in\Z} \ell\, T_\ell^{[M]}.\quad
\end{eqnarray}
Considering (\ref{sumrule1}) immediately follows (\ref{sumrule2}).

\noindent%
\emph{Proof of (\ref{sumrule3}).} For the compactly supported
scaling functions of Daubechies with 6 or more parameters the
function $x^2$ is still among the exactly expandable functions.
\begin{equation}\label{x2expansion}
  \sum_{\ell\in\Z} c^{[M]}_\ell s_{M\ell}(x) =x^2
\end{equation}
where the expansion coefficients are
\begin{equation}\label{x2expansioncoeff}
c^{[M]}_\ell=\int x^2 s_{M\ell}(x) dx=
 2^{-5M/2}(\mu_2+2\ell\mu_1+\ell^2),
\end{equation}
after similar steps applied in the previous proof. The notation
$\int y^2 s(y) dy=\mu_2$ was introduced. Differentiating
(\ref{x2expansion}) leads to
\begin{eqnarray}\label{proof3}
  \sum_{\ell\in\Z} c_\ell^{[M]}T_\ell^{[M]}&=&
  \int x\, s'_{M0}(x) dx=                                   \nonumber\\
  &&-\int s_{M0}(x) dx=-2^{-M/2}.\quad
\end{eqnarray}
The second equality follows from partial integration, and from the
fact that the integrated part disappears due to
$s_{M0}(\pm\infty)=0$. Substituting (\ref{x2expansioncoeff}) into
(\ref{proof3}) and using (\ref{sumrule1}) and (\ref{sumrule2})
gives (\ref{sumrule3}).

\end{document}